\documentclass[%
 reprint,
showkeys,
nobibnotes,
 amsmath,amssymb,
 aps,
]{revtex4-2}

\usepackage{graphicx}
\usepackage{amsmath} 
\usepackage{dcolumn}
\usepackage{bm}
\usepackage{hyperref}
\usepackage{tikz}
\usepackage{bm}
\usepackage{color}

\begin{document}

\title{Effective diffusion along the backbone of combs 
with finite-span 1D and 2D fingers}

\author{Giovanni Bettarini}
\email{giovanni.bettarini2@edu.unifi.it}
\affiliation{%
 Università di Firenze, Dipartimento di Fisica \& Astronomia, Via G. Sansone 1, 
 50019 Sesto Fiorentino, Italy
}%
\author{Francesco Piazza}%
 \email{Francesco.Piazza@unifi.it, corresponding author}
\affiliation{%
 Università di Firenze, Dipartimento di Fisica \& Astronomia and 
 INFN, sezione di Firenze, Via G. Sansone 1, 
 50019 Sesto Fiorentino, Italy
}%

\date{\today}
\begin{abstract}
\noindent
Diffusion in complex heterogeneous media such as biological tissues or porous materials 
typically involves constrained displacements in tortuous
structures and {\em sticky} environments. Therefore, diffusing particles experience
both entropic (excluded-volume) forces and the presence of 
complex energy landscapes. In this situation, one may describe transport through 
an effective diffusion coefficient. 
In this paper, we examine comb structures with 
finite-length 1D and finite-area 
2D fingers, which act as purely diffusive traps. 
We find  that there exists a critical 
width of 2D fingers above which the effective diffusion along the backbone is 
faster than for an equivalent arrangement of 1D fingers.
Moreover, we show that the effective diffusion coefficient
is described by a  general 
analytical form for both 1D and 2D fingers, 
provided the correct scaling variable is identified 
as a function of the structural parameters.  Interestingly, this formula 
corresponds to the well-known general situation 
of diffusion in a medium with fast reversible adsorption. Finally, we show
that the same formula describes diffusion in the 
presence of dilute potential energy traps, e.g. through a landscape of square wells.
While diffusion is ultimately always the results of microscopic interactions 
(with particles in the fluid, other solutes and the environment), effective representations
are often of great practical use. The results reported in this paper help 
clarify the microscopic origins and the applicability of global, integrated descriptions
of diffusion in complex media.
\end{abstract}
\keywords{Effective diffusion, trapping, combs with finite-length fingers} 
\maketitle


\section{Introduction\label{s:intro}}

%
\noindent The study of transport in heterogeneous and disordered structures~\cite{Torquato}
has a long history, with applications in a wide range of fields,
from porous systems, aerogels, glasses and biological media~\cite{Havlin:2002}.
In particular, molecular diffusion has been widely investigated in 
tortuous and porous structures~\cite{Jacob} through a large array of
theoretical tools~\cite{Tartakovsky:2019aa}. These range from the classic
Fick-Jacobs~\cite{jacobs1967diffusion,Zwanzig:1992aa,Traytak2013} and related 
approaches to effective medium strategies~\cite{Choy}, pioneered by J. C. Maxwell 
when he considered transport in a heterogeneous material made of 
a good conductor with interspersed insulating  inclusions,
leading to the classic Maxwell Garnett~\footnote{Note that James Clerk Maxwell 
Garnett (1880-1958), a namesake of the more famous James Clerk Maxwell, 
was intentionally named after his father's friend James Clerk Maxwell.} 
theory~\cite{maxwell:1904,Kalnin:1998}.\\
%
%
\indent A useful, simple but non-trivial model where the impact of disorder 
and tortuosity can be readily explored is offered by comb-like structures.
These are lattices consisting 
of a main 1D chain, which we refer 
to as the {\em backbone}, from which 
several side-chains branch off (fingers).
Of course, a difference should be made from the start between combs 
with infinite-length branches and combs with finite-length fingers. 
In the first case, one is often dealing with intrinsically fractal structures,
or systems where waiting times associated with the residence in infinite
sublattices that branch off the backbone diverge.  Such situations, which generalize 
to the case of graphs, often lead to anomalous 
diffusion~\cite{ralf2000,Havlin:2002,Burioni2003,
BurioniCassiVezzani2004,trifce2016}. Here we are explicitly concerned 
with comb structures with finite-length fingers, which display ordinary diffusion 
along the backbone as a direct consequence of the finite waiting times in the 
fingers~\cite{Balakrishnan1995,Balakrishnan1996}. 
In fact, it has been demonstrated~\cite{Arous2015} that, when 
the mean time spent on fingers has a finite expectation value,
the asymptotic transport regime is always Brownian motion~\footnote{In order 
to observe non-Markovian dynamics (i.e. anomalous diffusion with memory) 
as the scaling regime in the presence of finite-span 
teeth, a diverging waiting time on the teeth/traps has to be
explicitly enforced as an additional ingredient of the model~\cite{Arous2015}}. 
Of course, the possibility still exists, depending on the lengths and 
configurations of the fingers, that short-time and long-time diffusion 
be characterized by different diffusion coefficients. In a log-log 
plot of the mean square displacement (MSD) one would observe sublinear stretches 
that connect the two time-separated linear regimes. However, this apparent 
anomalous diffusion does not properly qualify as sub-diffusion (as e.g. on a infinite 
fractal support), similarly to lateral diffusion of proteins on the membrane 
of living cells, where the picket and fence structure imposed by different 
cytoskeletal elements identifies two well-separated regimes of normal diffusion~\cite{pickets2003}.\\
\begin{figure*}[ht!]
    \centering
    \includegraphics[width=\textwidth]{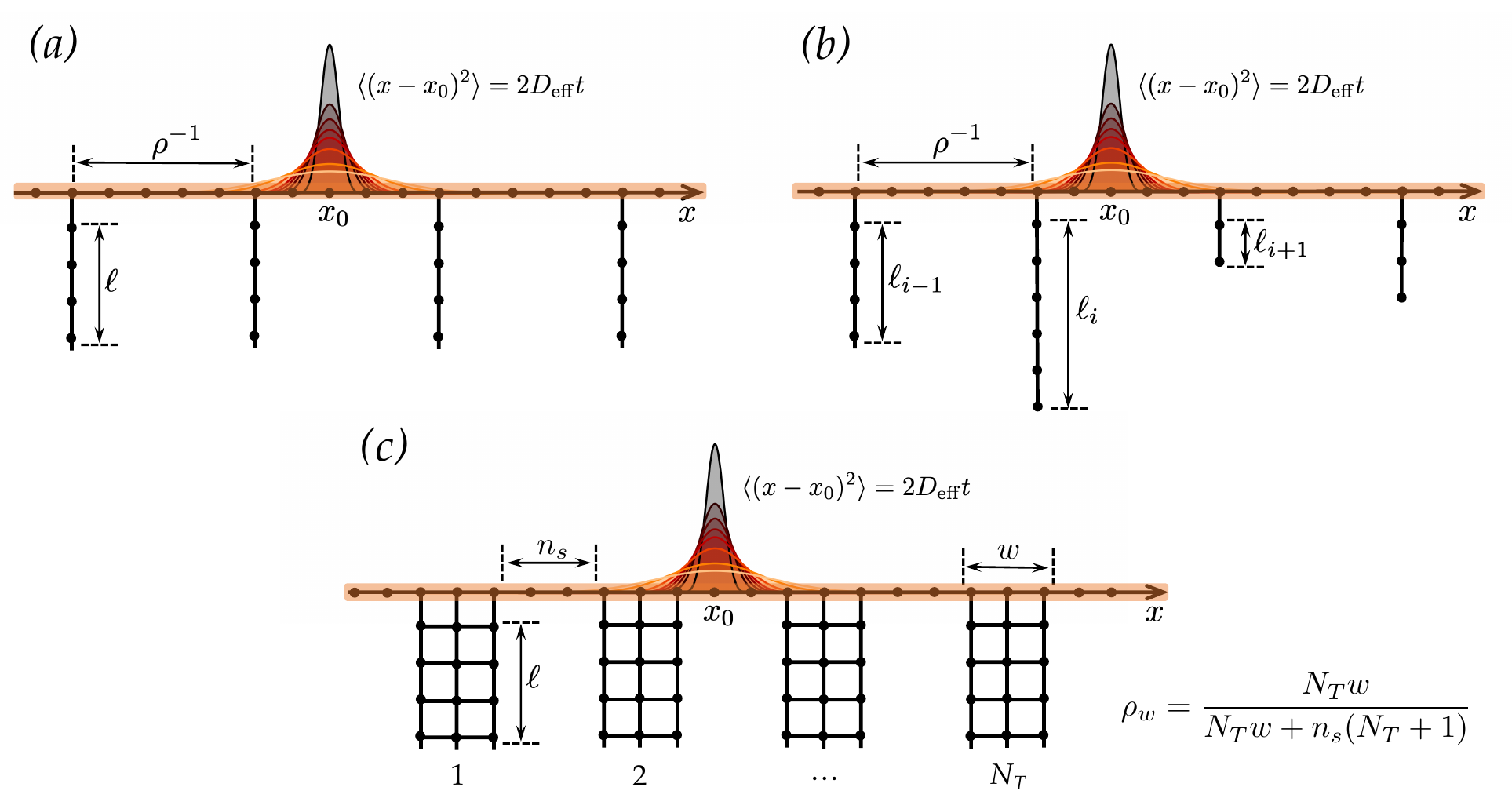}
   \caption{Illustration of the comb structures considered in this work. (a) 
            Regular configuration of $N_T$ identical traps of length $\ell$ branching away 
            at equally spaced sites of a chain comprising $N$ sites. (b) regular configuration
            of equally spaced traps of random lengths drawn from a given distribution 
            $\mathcal{P}(\ell)$ such that $\langle \ell\rangle < \infty$. In both cases the
            trap density is $\rho = N_T/N$. (c) Regular arrangement of 2D traps of fixed 
            width $w$ and length $\ell$. The density of trapping sites is $\rho_w$. The 
            horizontal shadings highlight the fact that we are only concerned 
            about diffusion along the {\em main} chain. The Series of Gaussian curves 
            illustrates our simulation protocol, where we launch a large number of 
            walkers from a given initial site $x_0$ (the backbone midpoint) 
            and compute the average variance of their 
            displacement as a function of time.}
    \label{fig:comb}
\end{figure*}
%
%
\indent In this paper we dwell on the properties of diffusion along the 
backbone of combs with finite-length fingers (see Fig.~\ref{fig:comb}).
More precisely, we investigate how the {\em effective} 
diffusion coefficient that describes 
diffusive spreading of trajectories along the backbone is influenced by the 
length and configuration of the fingers. In the following, we will 
turn to the word {\em trap} for all the structures that branch off the 
main backbone of the combs, as an explicit reminder that walkers that 
enter such structures are considered as {\em frozen} at their entry sites on the backbone 
for what concerns displacement along the latter. The paper is organized as follows. 
In section~\ref{s:1Dtraps} we describe our simulation 
protocol, which we first apply to examine diffusion along the 
backbone of combs decorated with different 
regular and random configurations of
1D finite-length fingers (traps). In  section~\ref{s:2Dtraps} we 
generalize to 2D traps, i.e. 2D-connected structures that branch off the backbone in 
correspondence of finite-length patches of consecutive branching nodes 
(see Fig.~\ref{fig:comb}). In section~\ref{s:disc}, we summarize our results 
and discuss them in the framework of an insightful 
analogy with problems where diffusion proceeds over an energy landscape, 
i.e in the presence of potential energy traps.

\section{Combs with 1D traps\label{s:1Dtraps}}

\noindent In all our simulations, many independent random walkers
are launched from the central site of the comb backbone, $x_0$.
Each walker performs a random jump process where at each time 
step $dt$ a new site is chosen randomly among the neighboring sites 
with a probability equal to the inverse of the connectivity. 
For example, for jumps in isolated stretches of the backbone
the jump probability is $1/2$ to the right and to the left.
If the walker happens to be at a site where a finger branches 
off, then one of its three neighbours is chosen randomly 
with probability $1/3$.
Walkers that enter a finger are considered as frozen at the 
entry node - hence the time advances as they diffuse in the 
finger, while their position on the comb backbone does not change. 
Typically, we compute the MSD $\mu_2(t)$ by 
considering how a large ensemble 
of $N_w$ walkers spread along the backbone, i.e. averaging over as many 
independet trajectories, that is 
\begin{equation}
    \label{e:mu2}
    \mu_2(t) = \frac{1}{N_w}\sum_{i=1}^{N_w} 
               \left(x_i(t) - x_0\right)^2
\end{equation}
where $x_i(t)$ is the coordinate of the $i-$th walker along the 
backbone at time $t$. In all the cases considered in this paper,
we obtained linear MSDs vs time (see also supplemental online 
material), from which an effective long-time diffusion 
coefficient could be measured unambiguously in all cases examined 
as $D_{\rm eff}= \lim_{t\to\infty}\langle \mu_2(t)/t \rangle/2$. The mean $\langle \dots \rangle$ 
denotes an average of $\mu_2(t)/t$ over many
time points well past possible transients.
Moreover, we systematically made sure that no walker ever reached 
the ends of the backbone at the end of the 
prescribed number of jumps.\\
%
\begin{figure*}[ht!]
    \centering
    \includegraphics[width=\columnwidth]{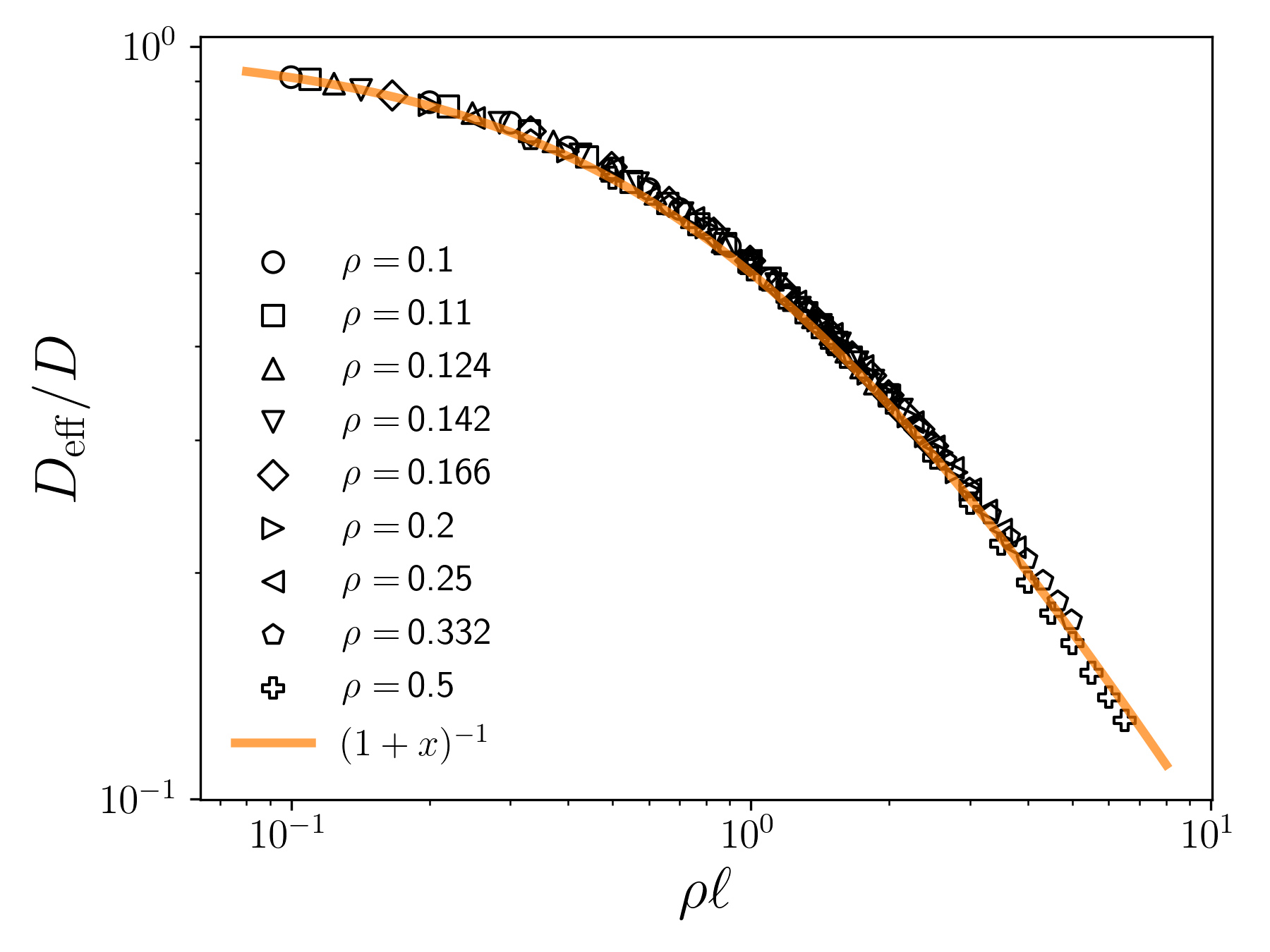}
     \includegraphics[width=\columnwidth]{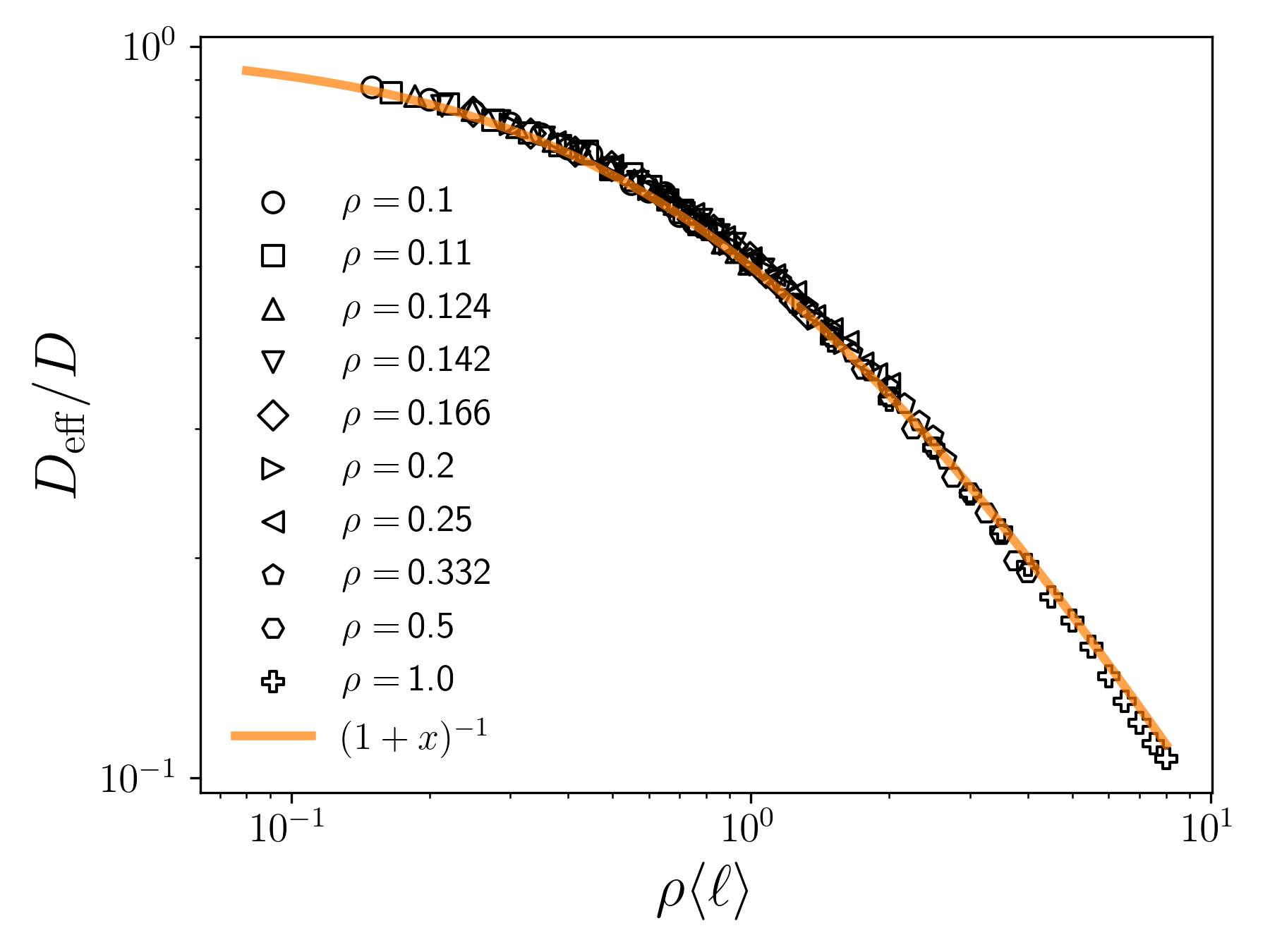}
    \caption{Normalized effective diffusion coefficient 
    measured in 1D comb-like lattices with regular branches/traps
    at fixed density $\rho$ (symbols). Here $D = a^2/2dt$.
    (a) Equal trap lengths $\ell\in [1,2,\dots,15]$,  
    (b) trap lengths uniformly distributed between $\ell_{\rm min}=1$ 
    and $\ell_{\rm max}=16$. For each pair $(\rho,\ell_{\rm max})$,
    $D_{\rm eff}$ has been averaged over 1000 independent realizations 
    of the $N_t=\rho N$ trap lengths. 
    The solid line is a plot of the simple 
    equilibrium prediction~\eqref{e:Deffx}.}
    \label{fig:Deff1D}
\end{figure*}
%
\indent In the first class of comb structures that we examined, 
both backbone and fingers are  one-dimensional lattices of sites 
with spacing $a$. All lengths in the following are expressed in 
units of $a$, and consequently linear density is expressed in units of $a^{-1}$.
We first analyzed configurations of $N_T$ traps 
(fingers) of equal length $\ell$ regularly spaced along a backbone 
comprising $N$ sites (Fig.~\ref{fig:comb} (a)). 
In this case the traps are equispaced at a distance $\rho^{-1}$, where  $\rho = N_T/N$ is their density. 
%
\begin{figure*}[ht!]
    \centering
    \includegraphics[width=\columnwidth]{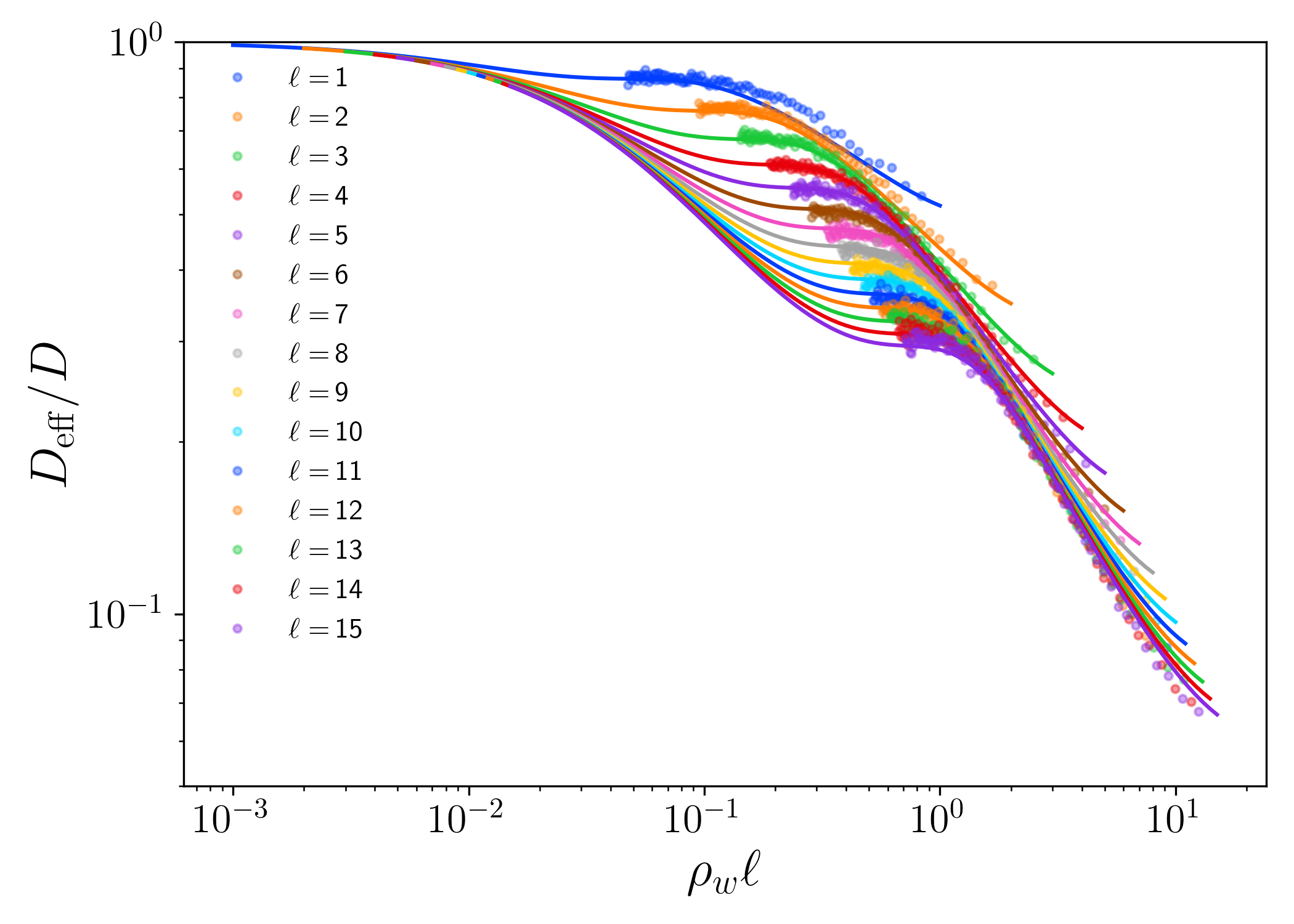}    
    \includegraphics[width=\columnwidth]{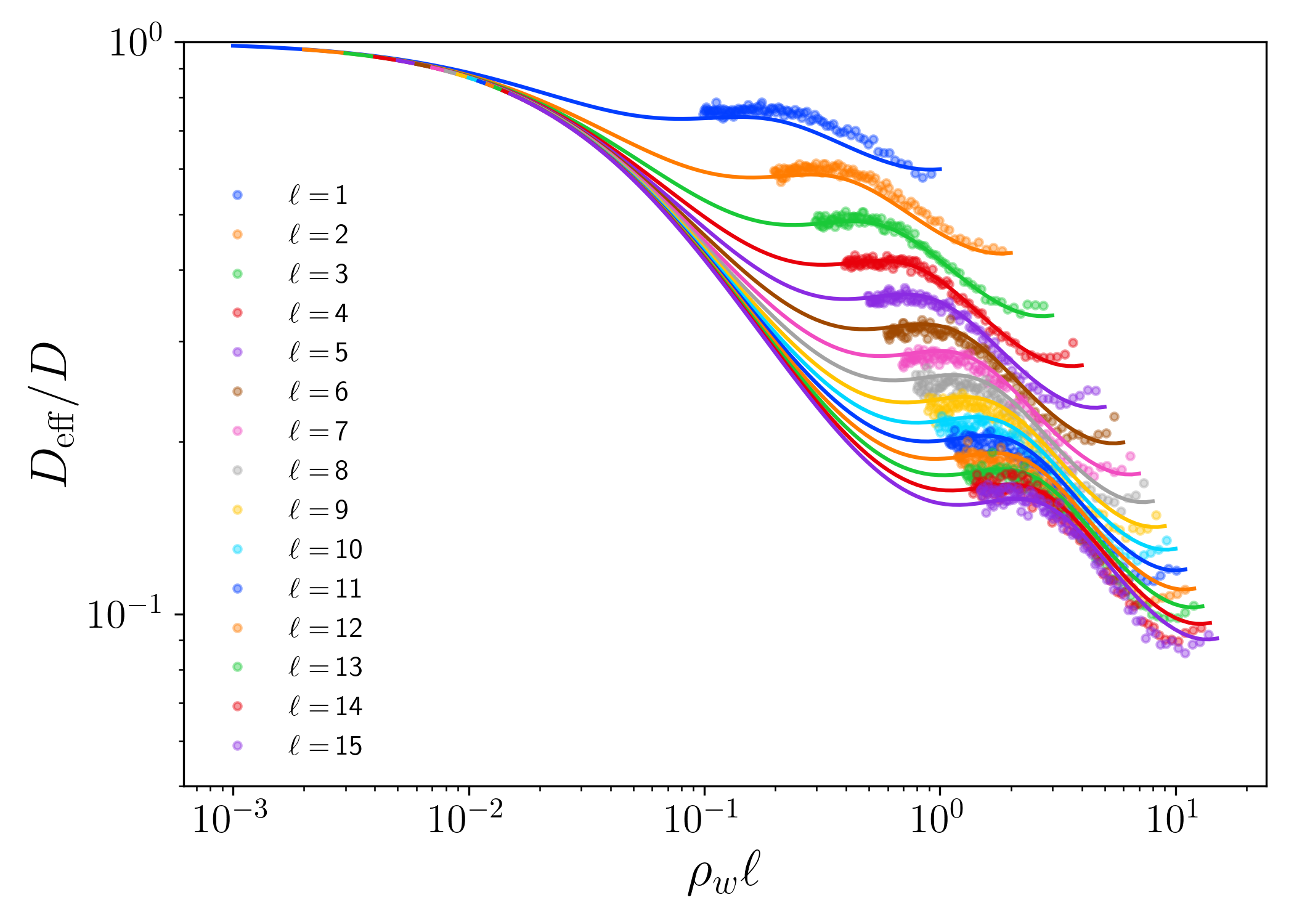}
    \caption{Normalized effective diffusion coefficient vs density 
             of trapping nodes on the backbone, $\rho_w$, for 
             different values of the traps' length, $\ell$.
             Left: $w=5$, Right: $w=11.$
             Symbols represent the simulation results, solid 
             lines are plots  of expression~\eqref{e:Deffw} corresponding 
             to the optimal function $\Gamma(\rho_w,w)$ (see text for 
             the fit details).}
    \label{fig:DnvsrhowallL}
\end{figure*}
The effective diffusion measured for a wide range of values 
of the parameters $\ell$ and $\rho$ is plotted in 
Fig.~\ref{fig:Deff1D} (left panel), 
normalized to the coefficient of 
free diffusion along the backbone, as a function of the 
non-dimensional variable $\xi = \rho \ell$. It  is apparent that 
all the different conditions lead to a value
of $D_{\rm eff}$ that is described by the same master curve in 
this reduced representation,
\begin{equation}
    \label{e:Deffx}
    \frac{D_{\rm eff}}{D} = \frac{1}{1 + \xi}
\end{equation}
In fact, when we considered configurations
of 1D fingers of random length (Fig.~\ref{fig:comb} (b)), 
we obtained identical results, provided the average trap length
is used to construct the scaling variable, 
i.e. $\xi=\rho\langle \ell \rangle$ (Fig.~\ref{fig:Deff1D}, right panel).
\noindent These results can be rationalized through a simple 
equilibrium argument.
Let us imagine a general situation where  
a particle is diffusing in a given 
finite volume $V$, bounded by reflecting walls. 
If we establish an arbitrary partition $V=V_1+V_2$, then at equilibrium
the proportion of time spent by the diffusing particle in each 
subregion $V_i$ $(i=1,2)$ is proportional to the corresponding 
volume fraction,  $\phi_i = V_i/V$~\cite{antipov2014}. 
Following a similar reasoning, our 1D comb-like lattice 
can be considered as a total (bounded) {\em volume} 
$V = (N + N_t\ell)\equiv V_c + V_t$, 
consisting of a main chain of volume 
$V_c=N$ and an overall trapping volume $V_t = N_t\ell$
(in units of the lattice spacing $a$).
During a given observation time $t$, a particle moves on the main 
chain only during the effective time $t_c = t V_c/(V_c + V_t)$.
Since the mean square displacement
is proportional to both the diffusion coefficient and
the time $t$, we can change from effective time to 
effective diffusion coefficient, so that 
\begin{equation}
    \label{e:Deff}
    D_{\rm eff} = D \frac{V_c}{V_c+V_t} 
                = D \frac{N}{N + N_t\ell} 
                = \frac{D}{1 + \rho\ell}
\end{equation}
where we have used the definition of the trap density $\rho = N_t/N$
(measured in units of $a^{-1}$).
Note that the prediction~\eqref{e:Deff} is expected to hold 
for regular comb-like lattices with branches of equal length $\ell$ 
equispaced at intervals of length $1/\rho$, as 
well as for random combs where the $N_t$ branches
depart from random sites of the main chain.\\
%
\begin{figure*}[t!]
    \centering
    \includegraphics[width=\textwidth]{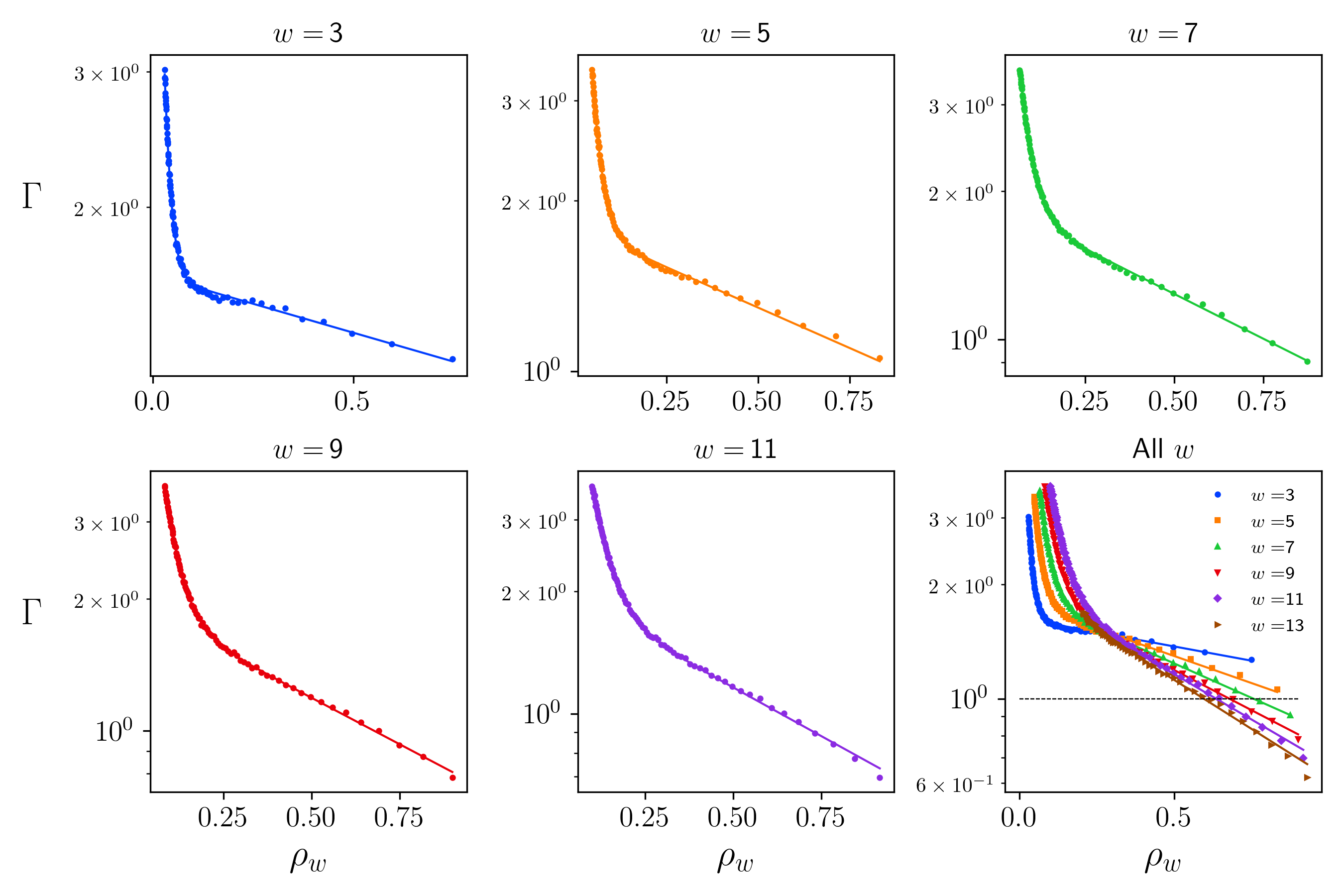}
    \caption{Functions $\Gamma$ obtained by fitting (solid lines) the effective 
    diffusion estimated from the simulations for all values of $\ell,N_T,w$ considered to $D_{\rm eff}=D/(1 + \Gamma \rho_w\ell)$. The  bottom 
    right panel is meant to illustrate the fact that $\Gamma$ crosses the 
    value 1 at lower and lower values of $\rho_w$ as the traps' width 
    $w$ increases (see also Fig.~\ref{fig:rhostar}).}
    \label{fig:Gammas}
\end{figure*}
%
\indent It is not difficult to see that the same argument should 
hold also when 
the traps are still equispaced but each trapping site is associated
with a trap of random length, drawn from a given normalized distribution 
$P(\ell)$ with finite mean. 
In this case, the overall trapping volume $V_t$  associated with 
the $N_t$ traps is obviously
\begin{equation}
    \label{e:trapVrand}
    V_t = \sum_{i=1}^{N_t} \ell_i = N_t\langle \ell\rangle
\end{equation}
where $\{\vec{\ell}\}$ is a given 
realization of $N_t$ random trap lengths. An argument 
identical to that followed for sets of equispaced but 
identical traps, coupled to Eq.~\eqref{e:trapVrand}, leads to 
\begin{equation}
    \label{e:DeffR}
    D_{\rm eff} = \frac{D}{1 + \rho\langle \ell \rangle}
\end{equation}
This result generalizes to arbitrary values of the trap density $\rho$
a known formula derived through 
mean first-passage time calculations in Ref.~\cite{Balakrishnan1996} for $\rho=1$.
Note that the expression~\eqref{e:DeffR} is expected to hold for a 
single arbitrary realization of a (sufficiently long) set of random trap lengths. Of course, it will hold {\em a fortiori} when averaged over many independent  realizations.  Moreover, as argued before 
for identical traps at regular intervals of 
length $1/\rho$, $N_t$ of them with random lengths {\em and} placed
at random with the same density $\rho$ will yield the same 
effective diffusion. 
Therefore, Eq.~\eqref{e:DeffR} is expected to describe a rather general situation, 
namely that of the effective
diffusion of a tracer in a 1D chain in the presence of diffusive
traps of random lengths departing from random locations along the backbone.

\begin{figure}[t!]
    \centering
    \includegraphics[width=\columnwidth]{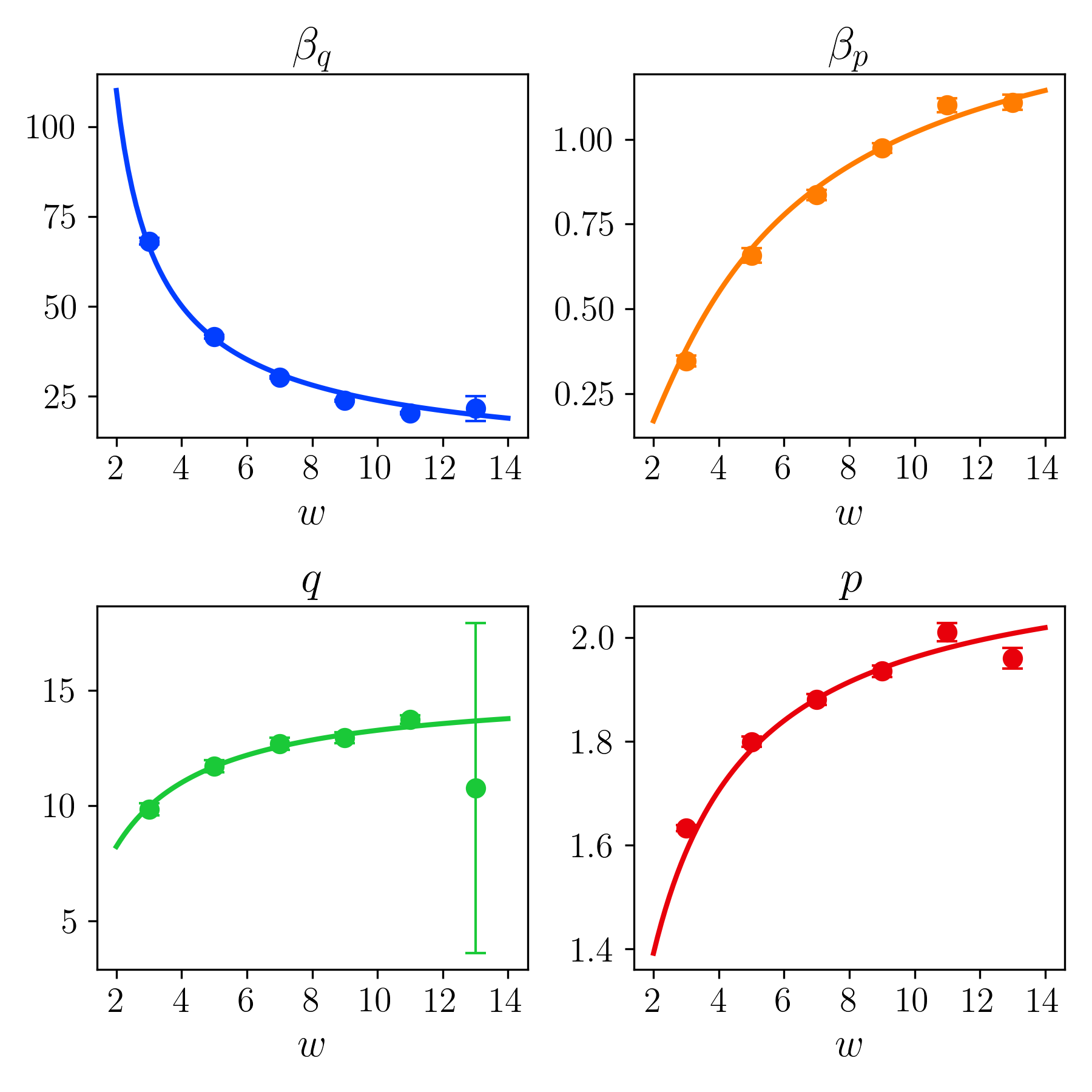}
    \caption{Best-fit values of the four parameters entering the 
    scaling function $\Gamma$, Eq.~\eqref{e:Gamma}, as functions 
    of the trap width $w$ (symbols). The solid lines are the best-fit 
    results of a constrained simultaneous interpolation of the 
    four quantities.
    The constrains are 
    $\lim_{w\to1}\Gamma(\rho_w(w),w)=1$,  
    $\lim_{w\to\infty}\Gamma(\rho_w(w),w)=1$
    (see the appendix for all the details).}
    \label{fig:Gammapars}
\end{figure}

\section{2D traps\label{s:2Dtraps}}

\begin{figure}[t!]
    \centering
    \includegraphics[width=\columnwidth]{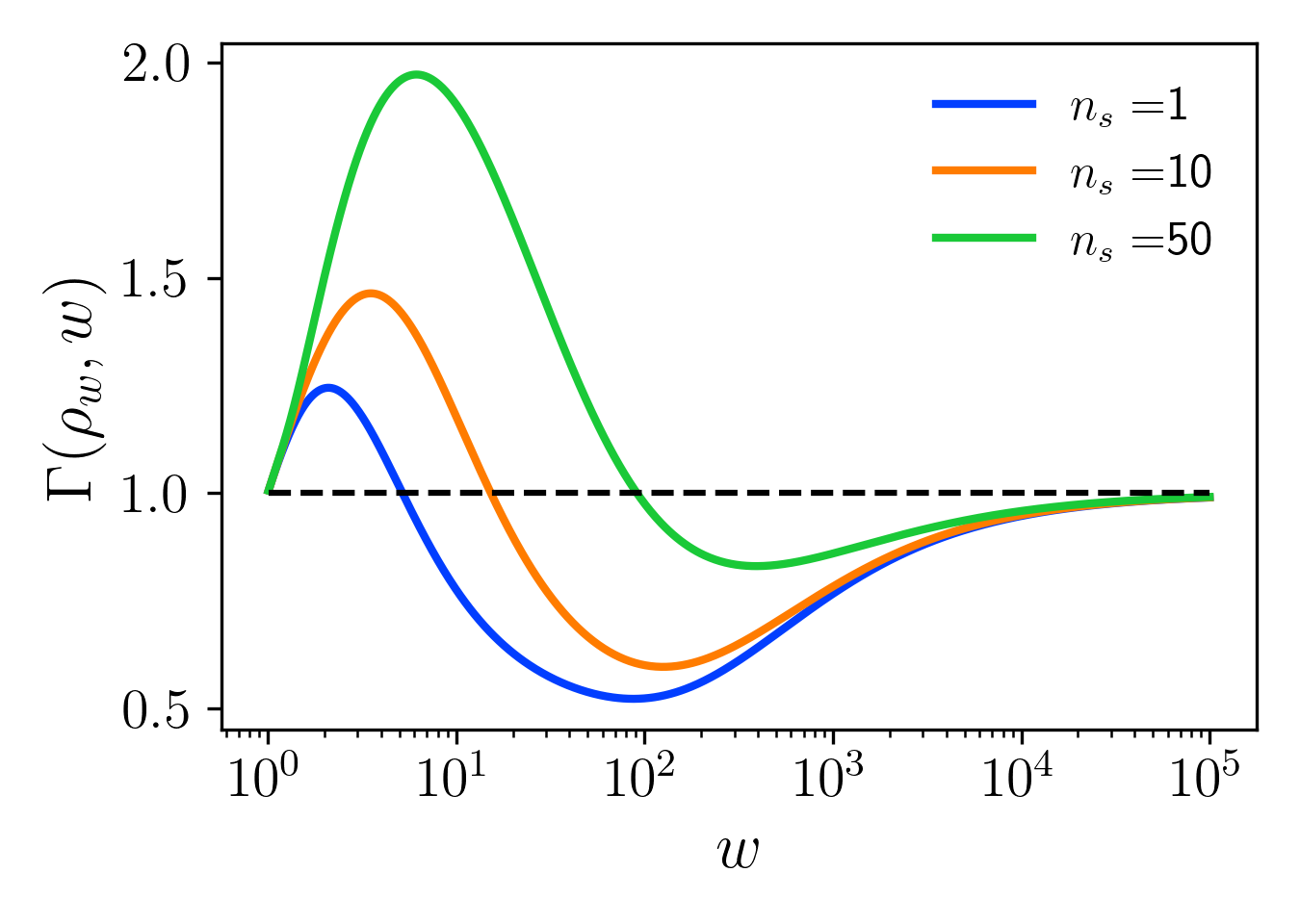}
    \caption{Plot of $\Gamma(\rho_w,w)$ over the whole 
    domain $w=(1,\dots)$  computed from the definition~\eqref{fig:Gamma} 
    with the best-fit interpolations of the functions 
    $q(w),p(w),\beta_q(w),\beta_p(w)$
    (see Table~\eqref{tab:Gammapars}) and in the limit $N_T\to\infty$, i.e. 
    with $\rho_w=w/(w+n_s)$.}
    \label{fig:Gamma}
\end{figure}

\noindent We have seen that in the presence of one-dimensional traps 
(i.e. comb fingers), the slowing down of effective diffusion on the backbone 
is captured by a simple effective volume argument. The scaling variable 
turns out to be the product of the trap length  $\ell$ (or the average 
length in the case of traps of different lengths) and the density of 
traps, $\rho$. As a consequence, effective diffusion on the 
comb backbone is independent of the configuration of traps. 
For example, so long as diffusion on fingers is one-dimensional, the effect of 
$N_T$ adjacent (i.e. clustered) or equispaced traps is exactly the same
(of course after averaging over many trajectories).
But what happens if identical fingers that depart at adjacent sites are 
connected by links at all sites ? 
In this case, traps become two-dimensional and the statistics of 
dwell (trapping) times is expected to change. Hence, one may wonder 
whether a simple effective volume argument will continue to hold. \\
\indent To address this question, in this section we investigate the simplest 
case of 2D traps, i.e. a backbone with $N_T$ 
identical 2D fingers of width $w$ and length $\ell$
equispaced at a distance $n_s$ (sites) along the backbone 
(see panel c in Fig.~\ref{fig:comb}).
In this case, the density of trapping sites generalizes to the 
following quantity
\begin{equation}
    \label{e:rhow}
    \rho_w = \frac{N_tw}{N_tw + n_s(N_T+1)}
\end{equation}
In accordance with a possible 
notion of {\em effective} diffusion along the backbone, 
we consider the $x$ coordinate of a walker as 
frozen once it has entered the 2D trap.
Time still ticks when the walker wanders in a 2D trap, but 
displacements are only measured on the comb backbone, 
i.e. at the time when 
(and at the site where) the walker will emerge again on the backbone  (not necessarily the same location 
as the entry site for $w \geq 2$). 
This situation would correspond to an observer that does not 
know that there are fingers and hence only {\em see} the backbone.  This observer would see the particle 
disappear from time to time, and appear again at different locations some time later.\\
\indent For $w=1$, i.e. when $N=(n_s+1)N_T +n_s$, $\rho_w$ 
indeed coincides with the density $\rho$ that describes combs with 
1D fingers. It should be noted that the definition~\eqref{e:rhow}
for $n_s=0$, although it represents a fully connected 2D strip of 
width $\ell$, still applies to the case where the walker cannot
hop from a finger to the next one in the {\em bulk} of the strip.
Hence, the case $n_s=0$ does not correspond to a fully connected 2D strip.
This limit is only meaningfully reached in our setting in the 
limit $w\to\infty$. \\
\indent One may think that $\rho_w$ could be used 
to construct a simple scaling variable that gauges the volume 
fraction of free diffusion as in the case of 1D traps. 
However, it can be seen immediately that in the case 
of 2D traps $\rho_w\ell$ does not allow to do so. 
In fact, as shown in Fig.~\ref{fig:DnvsrhowallL}, 
a plot of $D_{\rm eff}$ as a function 
of $\rho_w\ell $ does not yield a collapse as in the case of 
1D traps. However, we observe that the variables 
$\rho_w$ and $\ell$ can still be decoupled through a less
obvious mapping rule, $\rho \to \rho_w \Gamma(\rho_w,w)$.
This leads to  a generalization of Eq.~\eqref{e:Deff}  
to the following form
%
\begin{figure}[t!]
    \centering
    \includegraphics[width=\columnwidth]{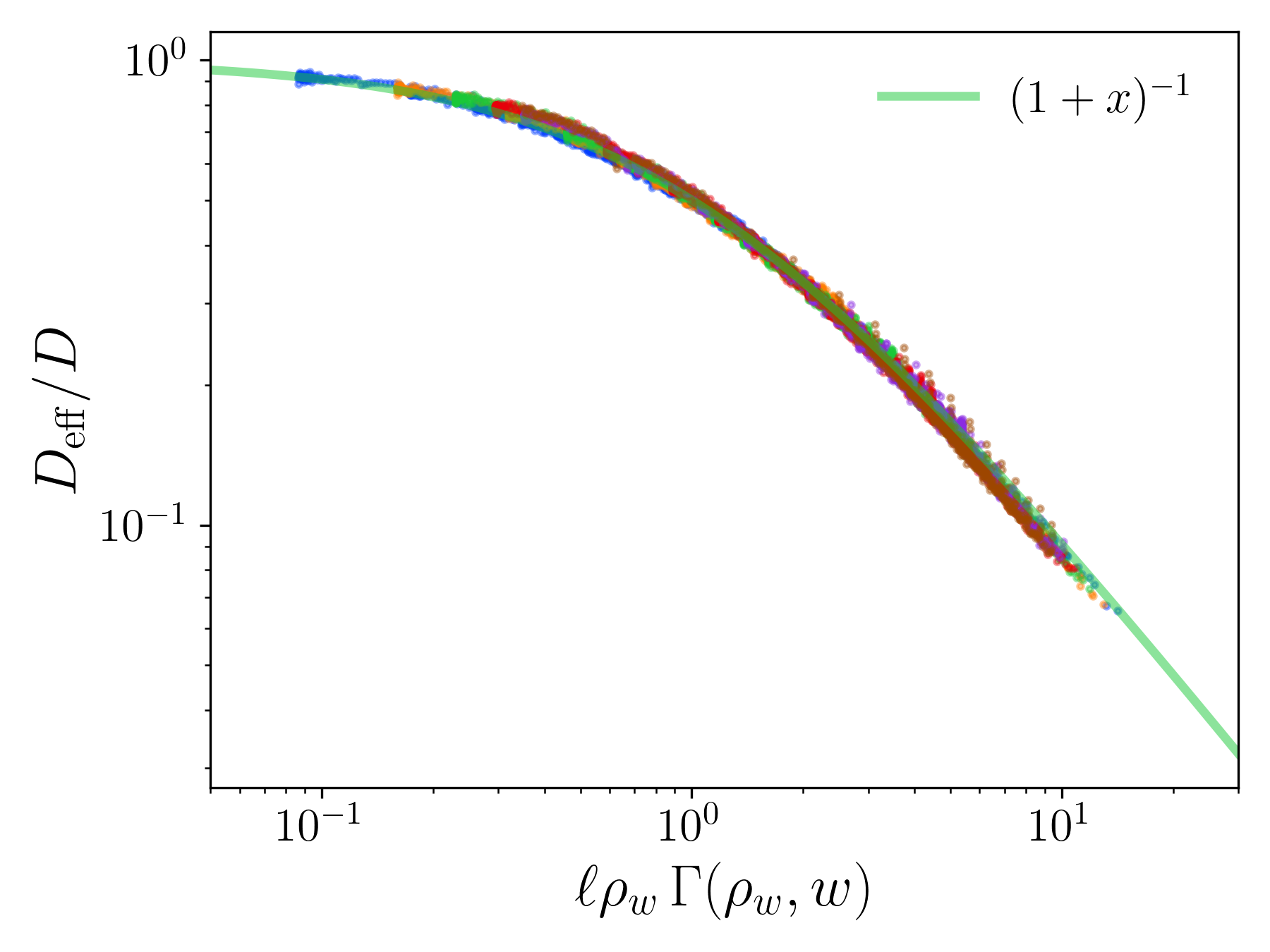}
    \caption{Effective backbone normalized diffusion constant for combs 
    featuring regularly
    spaced 2D traps of fixed width $w\in[3,5,7,9,11,13]$, 
    length $\ell\in[1,2,\dots,15]$ 
    and density $\rho_w$ as a function of the scaling variable 
    $\ell\rho_w \Gamma(\rho_w,w)$. Numerical 
    data collapse (symbols) and master curve (solid line).}
    \label{fig:rhowGcollapse}
\end{figure}
%
%
\begin{equation}
\label{e:Deffw}
D_{\rm eff}= \frac{D}{1 + \rho_w\Gamma(\rho_w,w)\ell}
\end{equation}
This phenomenology is illustrated in Fig.~\ref{fig:DnvsrhowallL}
for two different values of the traps' width $w$. 
Indeed, Fig.~\ref{fig:rhowGcollapse}  proves that the collective variable 
$\xi=\ell\rho_w\Gamma(\rho_w,w)$ leads to a data collapse  
corresponding to all the measured values of the 
diffusion coefficient along the backbone
on the same  master curve, namely $D_{\rm eff}/D= 1/(1+\xi)$, for 
a wide choice of values of the parameters $\ell$, $w$ and $n_s$.
An interesting implication of the particular analytic form 
of this recurrent master curve will be discussed in the 
final section of the paper. \\
\indent The function $\Gamma(\rho_w,w)$ is found to be a 
double exponential of the form
\begin{equation}
\label{e:Gamma}
\Gamma(\rho_w,w) = q(w) e^{-\beta_q(w) \rho_w} +
                 p(w) e^{-\beta_p(w) \rho_w}
\end{equation}
where the variables $\rho_w$ and $w$ appear decoupled. 
The parameters $q(w)$, $p(w)$, $\beta_q(w)$ and $\beta_p(w)$
in Eq.~\eqref{e:Gamma} are fitting parameters that depend 
on the trap width $w$.
From an operative point of view, we have uncovered this by first fitting 
the effective diffusion coefficient to a generic function of the 
form $D_{\rm eff} = D/(1 + \Gamma \rho_w\ell)$. We then realized 
that the quantity  $\Gamma$ does not depend on the traps' length 
$\ell$, but only depends on the parameters that specify
their density, $N_T,n_s$, through their combination $\rho_w$
and on their width, $w$. A plot of $\Gamma$ as a function 
of $\rho_w$ for a fixed $w$ makes it clear that it is a double exponential of 
the form~\eqref{e:Gamma}.  This is illustrated in 
Fig.~\ref{fig:Gammas}. In this case, $\rho_w$ is varied by 
changing the trap spacing $n_s$.\\
\indent The dependence of the best-fit parameters
on the traps' width is illustrated in Fig.~\ref{fig:Gammapars}. 
It appears that a 
limit asymptotic scaling function $\Gamma$ may be reached as 
$w$ increases. A simple argument can be invoked (see Appendix)
to demonstrate  
that the expected behaviour is
\begin{equation}
    \label{e:gammalim}
    \lim_{w\to\infty} \Gamma(\rho_w,w) = 
    q^\infty e^{-\beta_q^\infty} +
                 p^\infty e^{-\beta_p^\infty} = 1
\end{equation}
where
\begin{equation}
\label{e:qplim}
[q^\infty,p^\infty,\beta_q^\infty,\beta_p^\infty]=
\lim_{w\to\infty} [q(w),p(w),\beta_q(w),\beta_p(w)]    
\end{equation}
and we have used the definition~\eqref{e:rhow} to set 
$\lim_{w\to\infty} \rho_w(w)=1$. Moreover, in the limit $w\to1$
combs with 2D fingers reduce to ordinary combs with single-chain
fingers with density $\rho$, 
which also means that ($\lim_{w\to1}\rho_w(w) =\rho$)
\begin{equation}
\label{e:Gammalim0}
    \lim_{w\to1} \Gamma(\rho_w,w) = 1  
\end{equation}
The solid lines in Fig.~\ref{fig:Gammapars} are the result of simultaneous fits 
of the four functions $q(w),p(w),\beta_q(w),\beta_p(w)$ 
with simple scaling forms with the constraints~\eqref{e:gammalim}
and~\eqref{e:Gammalim0} enforced (see Appendix for the 
mathematical details). The fact that an excellent simultaneous fit 
of the four parameters that respects the two global 
constraints is possible at all is a strong gauge of 
self-consistency of our interpretation. \\
\indent It is instructive to give the function $\Gamma$
a closer examination. For simplicity, we shall do so 
in the {\em thermodynamic} limit (infinite traps),
$N_T\to\infty$, so that $\rho_w = w/(w + n_s)$. The function 
$\Gamma(w,n_s)$ is plotted in Fig.~\ref{fig:Gamma} for different 
values of the trap spacing $n_s$ (see again Fig.~\ref{fig:comb}).
It may be noticed that $\Gamma$ becomes lower than 1
at a specific trap width $w^\ast$ that increases with $n_s$, 
reaches a minimum and saturates to 1 again as it should, given 
the constraint~\eqref{e:gammalim}. Overall, the important 
message here is that 
$\Gamma(w,n_s) < 1$ for $w>w^\ast(n_s)$.\\
\indent A value of $\Gamma$ less than unity may be interpreted as 
describing a situation 
where the diffusion along the backbone in the presence of 2D traps 
is less hindered (proceeds
faster) than an equivalent arrangement of 1D traps of equal length
with a density $\rho$ equal to the 2D density 
$\rho_w$.
In order to clarify the meaning of this finding, let us consider 
a given number $\rho N$ of 1D fingers of length $\ell$ on a backbone 
comprising $N$ sites, and let us imagine 
to arrange them in {\em chunks} of $w$ adjacent fingers. 
In this situation we have seen (independently of the 
configuration) that the effective diffusion is $D_{\rm eff}=D/(1 + \rho\ell)$.
Let us imagine a second scenario, where the configuration is identical 
but we have {\em switched on} connections among the sites 
that belong to adjacent fingers, so that these become 2D traps of width $w$.
Keeping in mind that the trap density is the same in both 
cases, $\rho_w=\rho=w/(w+n_s)$, we conclude that
below $w^\ast(n_s)$ a number $w$ of adjacent but unconnected 
1D fingers yield slower diffusion along the backbone
than if these were full 2D traps of width $w$. 
Conversely, for $w>w^\ast(n_s)$ diffusion along the backbone
proceeds faster for the full 2D traps.\\
%
%
\indent The critical trap width $w^\ast$ increases with the 
spacing $n_s$. A simple analysis shows that 
\begin{equation}
    \label{e:wstar}
    \begin{cases}
        [w^\ast] = n_s + 5 & \text{for} \ 1\leq n_s \leq 6 \\
        [w^\ast] = n_s + 6 & \text{for} \ 6< n_s \leq 16\\
        [w^\ast] = n_s + 7 & \text{for} \ 16< n_s \leq 21\\
        [w^\ast] = n_s + 8 & \text{for} \ 21< n_s \leq 24\\
        \dots & \\
        [w^\ast] \propto n_s^4 & \text{for} \ n_s \gg 1
    \end{cases}
\end{equation}
where $[\dots]$ denotes the nearest larger integer. Furthermore, 
we note that the {\em acceleration} factor (represented by the 
value of $\Gamma$ at its minimum) lies
between 0.5 and 1 (i.e. no difference) for $n_s\in[1,\infty)$
(see again Fig.~\ref{fig:Gamma}).\\
\indent Intuitively, this curious fact 
might be explained by the fact that, as the entry region of the 
traps along the backbone increases, so does the probability that 
a walker entered somewhere will leave the trapping volume in correspondence
of a non-adjacent junction, as opposed to the situation where to 
get to the same junction the walker could wander in each
isolated 1D finger in between. \\
\indent Finally, one may wonder whether it is the very definition of effective 
diffusion that we used for 2D traps that lies at the heart of the phenomenology 
illustrated in Fig.~\ref{fig:rhowGcollapse}. 
Another (maybe more intuitive) definition of effective diffusion 
along the backbone would be to simply count {\em all} displacements along $x$, 
whether the walker is within a 2D finger or in a trap-free region along 
the backbone. The results of calculations performed by following 
this method are identical to those illustrated in this section (see 
supplemental online material).

\section{Summary and discussion \label{s:disc}}

\noindent In this work we have investigated the effective diffusion along 
a 1D chain decorated with different configurations of 1D and 2D diffusive 
traps. Walkers that enter a trap from a junction site are considered as 
frozen at that position for what concerns displacements along the backbone, 
until  they get back to it (at the same or at a different site).
Therefore, the effective diffusion gauges transport as it would be described
by an observer oblivious of everything but traveling legs along 
the backbone. In the spirit of continuous time
random walks, such observer would describe diffusion as sequences of jumps
separated by waiting times, whose statistics would appear to incorporate unusually
long times. These would correspond to time stretches where walkers wander 
in the trapping volume but do not contribute to transport along the backbone. \\
\indent We found that the average diffusive spreading of many trajectories
is well captured by a simple equilibrium argument in the case of 1D fingers,
even if averages are performed exclusively on non-equilibrium trajectories
(we have always made sure that none of them had reached the ends of the backbone).
In the case of 2D traps, i.e. fully connected adjacent 1D fingers,
the simple equilibrium argument breaks down. However, we found that 
the effective diffusion coefficient (in units of the 
free diffusion along  the backbone) is described by the same analytical form
as for isolated 1D traps, namely $1/(1+\xi)$, irrespective of the 
definition of $D_{\rm eff}$, provided the correct 
scaling variable $\xi$ is identified. While in the case of 1D 
finite-length fingers one simply has $\xi = \rho\langle\ell\rangle$, where 
$\rho$ is the density of fingers and $\langle\ell\rangle$ their average 
length, in the case of 2D traps, the scaling variable $\xi$ takes a more 
complicated form as a function of the parameters that define 
the trapping volume (see Eq.~\eqref{e:Deffw}). However, 
the master curve is still the same, so that one may wonder whether there is 
some deep physical meaning to it.\\
\indent In order to shed more light on this observation, 
it is instructive to draw an analogy with diffusion in the presence of 
potential energy traps, e.g. binding sites.
Let us start with a simple text-book kinetic argument in the continuum 
limit (see e.g. Ref.~\cite{Crank}).
Let us consider diffusion in 1D in the presence of a certain 
density $\rho$ of {\em sticky} sites, where diffusive particles can be immobilized with 
rate $k_{\rm on}\rho$ and from where they can detach with rate $k_{\rm off}$.
If $c$ denotes the concentration of freely diffusing particles and $s$ that 
of immobilized ones, the evolution of  the field 
$c(x,t)$ is governed by the following equation
\begin{equation}
    \label{e:diffcs}
    \frac{\partial c}{\partial t} = D\frac{\partial^2 c}{\partial x^2} 
                   - \frac{\partial s}{\partial t}     
\end{equation}
In the hypothesis that the binding and unbinding kinetics is faster than 
free diffusion, one may formulate a local equilibrium hypothesis, namely
$s \propto c$. A simple physically meaningful choice would be to take 
$s = (k_{\rm on}\rho/k_{\rm off}) c \equiv (\rho/\mathbb{K}) c$, where 
$\mathbb{K}$ is the equilibrium (dissociation) constant that describes 
the binding process. The consequence of this choice is that Eq.~\eqref{e:diffcs} 
is mapped to a transport equation where diffusion is described by the modified 
diffusion coefficient
\begin{equation}
    \label{e:Diffmod}
    D_{\rm eff} = \frac{D}{1 + \rho/\mathbb{K}}
\end{equation}
Comparing Eq.~\eqref{e:Diffmod} with Eqs.~\eqref{e:Deff} and~\eqref{e:DeffR}, one
is led to
conclude that diffusive trapping is equivalent to diffusion in the 
presence of fast binding/unbinding
dynamics with a density of immobilization sites identical to that of 
finite-length fingers. Then, the association constant that gauges 
the binding strength (i.e. $\mathbb{K}^{-1}$) plays 
the role of the  average length of diffusive traps (i.e. the fingers) - 
the longer the traps/the lower the dissociation constant $\mathbb{K}$,
the stronger the bond/the trapping power.\\
\indent We are now ready to examine how trapping walkers 
in potential energy wells compares with diffusive trapping in 
comb fingers. Let us imagine a 1D lattice where narrow square 
wells of depth $\epsilon$ are present with density $\rho$.
In order to compare with 1D fingers, 
we imagine the wells to have a range equal to one lattice spacing $a$
(our coarse-grained length unit). Potential wells will be then 
separated by an average distance $d=\rho^{-1}$ (in units of $a$).
Taking a {\em module} to be between $0$ and $d$,
we can take the well at a point $x_0$ anywhere within the 
interval, namely
\begin{equation}
    \label{e:pot}
    U(x) =
    \begin{cases}
        0           & 0 \leq x < x_0 \\
        -\epsilon   & x_0 \leq x < x_0 + a \\
        0           & x_0 + a \leq x \leq d\\
    \end{cases}
\end{equation}
The 1D lattice can be thought of as characterized by an energy 
landscape comprising a repetition of identical modules
given by the energy landscape~\eqref{e:pot} and corresponding 
to a given sequence of (in general different) 
$x_0$ values. The effective diffusion coefficient 
for such a system can be computed through a well-known 
formula that describes diffusion in periodic energy
landscapes~\cite{festa1978},
\begin{equation}
    \label{e:festa}
    D_{\rm eff} = \frac{D}{\langle e^{\beta U} \rangle
                           \langle e^{-\beta U}\rangle}
\end{equation}
where $\langle \dots \rangle$ denotes an average over one period. 
In our case, we have 
\begin{equation}
    \label{e:averU}
    \langle e^{\pm\beta U} \rangle = 
    \frac{1}{d}\int_0^d e^{\pm\beta U(x)}dx =  
                        1 + \rho \left( e^{\mp\beta \epsilon}-1\right)
\end{equation}
where $\rho$ is measured as in the case of combs in units of $a^{-1}$.
In this case, we can interpret $\rho$ as the finite limit of the 
quantity $N_T/(Na)$ as $N,N_T\to\infty, a\to 0$.
We note that these averages do not depend on the specific locations
of the potential wells, $x_0$, so that the overall situation is indistinguishable 
from a periodic energy landscape. 
Plugging Eq.~\eqref{e:averU} in formula~\eqref{e:festa}, 
we readily obtain 
\begin{equation}
    \label{e:DeffU}
    D_{\rm eff} = \frac{D}{1 + 2\rho(1-\rho)(\cosh \beta\epsilon -1)}
\end{equation}
It can be appreciated that Eq.~\eqref{e:DeffU} belongs to 
the same general form as our previous results, with a 
scaling variable that appears nonlinear in $\rho$.
Note that Eq.~\eqref{e:DeffU} predicts correctly that $D_{\rm eff}=D$ for 
$\rho = 0$, but also for $\rho=1$. In fact, in the latter case 
the potential is flat everywhere, so that diffusion proceeds 
freely as it should. Hence, the regime where one should match
the result~\eqref{e:DeffU}  with its analogue for combs with 
1D fingers is the linear regime, $\rho \ll 1$. In this case, we have
\begin{equation}
    \label{e:DiffU12}
    \frac{D_{\rm eff}}{D} \simeq
    \begin{cases}
        \left[1 + \rho (\beta \epsilon)^2 \right]^{-1}   & \beta\epsilon \ll 1 \\
         \left[1 + \rho \,e^{\beta \epsilon} \right]^{-1}  & \beta\epsilon \gg 1 
    \end{cases}
\end{equation}
Inspection of the two limiting behaviours~\eqref{e:DiffU12} is instructive. 
For shallow potentials wells, the analogue of the average finger length 
appears to be $(\beta\epsilon)^2$. Incidentally, we note that this agrees with  
a well-known result derived by Zwanzig for diffusion in spatially rough 
energy landscapes~\cite{Zwanzig1988}, 
as $\left[1 + \rho (\beta \epsilon)^2 \right]^{-1} 
\simeq \exp [-\rho (\beta \epsilon)^2]$ in the regime $\beta\epsilon\ll 1$. \\
\indent In the opposite limit of deep energy wells, we recover a 
physically intuitive result. In this case, the effective diffusion 
takes the following rather simple form,
\begin{equation}
    \label{e:DeffUArr}
    D_{\rm eff} = \frac{D}{1 + \rho \,e^{\beta\epsilon}}
\end{equation}
The analogy with diffusive trapping in 1D fingers 
is therefore $\ell \simeq e^{\beta\epsilon}$. More interestingly, 
and  physically reassuringly in its 
self-consistency, we note that in this regime 
one expects the escape out of the wells to be a thermally activated 
process. Hence, seen from the perspective of a walker 
that jumps 
over the energy landscape, once it has entered a potential 
well, the exit rate $k_{\rm off}$ should follow Arrhenius law, 
$k_{\rm off} \propto e^{-\beta\epsilon}$. Recalling Eq.~\eqref{e:Diffmod},
we conclude that $\mathbb{K} = e^{-\beta\epsilon}$. One may 
recognize the standard textbook definition of the affinity in 
terms of the binding energy of the complex.\\
\indent In conclusion, while diffusion fundamentally arises from 
microscopic interactions involving particles within the fluid, 
other solutes, and the surrounding environment, concise effective descriptions 
prove highly practical. The findings presented in this paper contribute 
to elucidate the underlying microscopic mechanisms and the 
utility of employing effective descriptive frameworks.

\section*{Supplementary Material}

\noindent The online supplemental material contains example 
plots of the MSD vs time in the absence and presence of traps (averaged 
over many independent trajectories) for several choices of the
trap density, width and length. Moreover, it contains the results of 
the simulations performed in combs with 2D traps 
by counting all displacements along $x$, whether the walker is within a 
finger or in a trap-free region along the backbone.

%
\appendix*  
\setcounter{equation}{0}  
\section{A\label{a:1}}

\noindent In this appendix we provide the mathematical details underlying the 
simultaneous fits of the four functions $q(w)$, $p(w)$, $\beta_q(w)$ 
and $\beta_p(w)$ shown in Fig.~\ref{fig:Gammapars}. We first  looked
for the simplest empirical scaling forms that could fit the functions 
separately one by one. However, as discussed in the main text, 
there exist two global constraints on the function $\Gamma(\rho_w,w)$ that 
should be respected by whatever combination of free parameters chosen 
to represent the functions $q(w)$, $p(w)$, $\beta_q(w)$ and $\beta_p(w)$. These read
\begin{eqnarray}
    &&\lim_{w\to 1} \Gamma(\rho_w(w),w) = 1 \label{e:limG1}  \\
    &&\lim_{w\to \infty} \Gamma(\rho_w(w),w) = 
                         \Gamma^\infty \label{e:limGi} < \infty
\end{eqnarray}
where we have made it explicit that $\rho_w$ is also a function of the trap 
width $w$. In particular, one has 
\begin{eqnarray}
    &&\lim_{w\to 1} \rho_w(w) = \rho \label{e:limrhow1}\\
    &&\lim_{w\to \infty} \rho_w(w) = 1  \label{e:limrhowinf}
\end{eqnarray}
where $\rho$ is the density of 1D fingers. \\
\indent A simple argument allows to show  that  
$\Gamma^\infty=1$.
Let us imagine a walker that hops with rate $k_D$ on a 1D lattice 
of spacing $a$, making right or left jumps with equal probability 
at each time step $dt$. The diffusion coefficient in this situation is 
$D=k_Da^2$.
Now, let us imagine that the walker can occupy 
$M$ different {\em chemical} states, and that hopping proceed at different 
rates in different chemical states, $k_D^\alpha$, with 
$\alpha=1,2,\dots,M$. This amounts to assume as many different 
diffusion coefficients as there are chemical states, 
$D_\alpha=k_D^\alpha a^2$. To picture a physical situation that 
may be described by this model, one can imagine a large non-globular protein 
that fluctuates among different conformational states,
each corresponding to a different hydrodynamic radius. 
We assume that transitions in {\em chemical space}
correspond to a Markov process and  can therefore 
be described by an associated master equation
with transition rate matrix $W$ (independent of space)
\begin{equation}
    \label{e:Wtrans}
    W_{\alpha\beta} = k_{\alpha\beta} - \delta_{\alpha\beta} 
                      \sum_{\sigma=1}^M k_{\sigma\alpha}
\end{equation}
where $k_{\alpha\beta}$ is the transition rate for the 
$|\beta \rangle \to |\alpha\rangle$ transition. 
Overall, the system is described by a coupled reaction-diffusion 
system for the set of probability densities $P_\alpha(x,t)$ of 
the form 
\begin{equation}
    \label{e:RDsys}
    \frac{\partial P_\alpha(x,t)}{\partial t} =
    D_\alpha \frac{\partial^2 P_\alpha(x,t)}{\partial x^2} + 
    \sum_{\beta=1}^M W_{\alpha\beta}P_\beta(x,t)
\end{equation}
Let us indicate with $\overline{p}_\alpha=\lim_{t\to\infty}p_\alpha(t)$ 
the equilibrium steady-state values of the occupancy probabilities 
in chemical space, i.e. the normalized null space of the transition 
rate matrix $W$ ($W \overline{\bm{p}} =0$, with 
$\sum_\alpha \overline{p}_\alpha=1$). The 
overall mean square displacement $\mu_2(t)$ of the walker is 
obtained by summing over all chemical states, namely
\begin{equation}
    \label{e:mu2}
    \mu_2(t) = \int_{-\infty}^\infty x^2 
               \sum_{\alpha=1}^M P_\alpha(x,t)\,dx
\end{equation}
It is not difficult to prove~\cite{Liang:2022} that the long-time 
diffusion coefficient of the walker, defined as 
$D = \lim_{t\to\infty} \mu_2(t)/2t$, is the equilibrium average 
of the single diffusion coefficients associated with the 
different chemical states, namely 
\begin{equation}
    \label{e:Dinfty}
    D = \sum_{\alpha=1}^M \overline{p}_\alpha D_\alpha
\end{equation}
This can be seen for example through a multiple scale 
expansion, where the role of the fast time is played
by the inverse of the typical rate that governs
transitions among chemical states. \\
\indent Let us now imagine that the different chemical 
states are replicas of the same 1D lattice, so that 
$k_D^\alpha=k_D$ $\forall \ \alpha$. 
In this case, our 2D chemical-spatial 
system is a full 2D spatial system. This also 
means that $k_{\alpha\beta} = k_D$ $\forall \ \alpha,\beta$.
More specifically, the system is now an infinite strip 
in the $x$ direction of width $M = \ell+1$.
In this case, one has  $\overline{p} = 1/M= 1/(1 + \ell)$.
Therefore, Eq.~\eqref{e:Dinfty} predicts that 
the overall diffusion coefficient of 
the strip in the $x$ direction is the same as 
along each individual 1D track, i.e. $D = k_D a^2$.\\
\indent Let us now come back to the limit $w\to\infty$ of 
the diffusion along the backbone of 2D fingers of length 
$\ell$ with density $\rho_w$. 
It should be clear now that this is 
exactly the case of the infinite 2D strip of width 
$M=\ell +1$. Therefore, the effective diffusion along 
the backbone in our case can  be identified by 
computing the mean square displacement along the bottom 
edge of the infinite strip, namely
\begin{equation}
    \label{e:Deffwinf}
    \mu^0_2(t) = \int_{-\infty}^\infty x^2 P_0(x,t)\,dx 
               = 2 D_{\rm eff} t
\end{equation}
With $k_D^\alpha=k_{\alpha\beta}=k_D$ $\forall \ \alpha,\beta$,
one has $\overline{p}_\alpha=1/(1+\ell)$ $\forall \ \alpha$.
Hence, from Eq.~\eqref{e:Dinfty} we get ($\alpha=0$ labels
the bottom edge of the strip)
\begin{equation}
     D_{\rm eff} = D\,\overline{p}_0 = \frac{D}{1 + \ell}
\end{equation}
Recalling that $\lim_{w\to\infty} \rho_w(w) = 1$, 
and comparing with Eq.~\eqref{e:Deffw}, we 
finally conclude that 
$\Gamma^\infty=1$ (see Eq.~\eqref{e:limrhowinf}).\\
\indent 
Equipped with the two constraints~\eqref{e:limG1}
and~\eqref{e:limGi} with $\Gamma^\infty=1$,
we can set out to find coherent analytical interpolations 
of the functions $q(w),p(w),\beta_q(w),\beta_p(w)$.
Based on the single-function examination conducted beforehand, 
we found that a simple choice of scaling functions 
that fulfills  the global constraints~\eqref{e:limG1} 
and~\eqref{e:limGi} reads
\begin{eqnarray}
    &&q(w) = q_0 + (q^\infty - q_0) 
                \left( \frac{w-1}{w+1} \right)  \label{e:q}      \\
    &&p(w) = 1 + (p^\infty - 1) 
                \left( \frac{w-1}{w+1} \right)  \label{e:p}      \\            
    &&\beta_q(w) = \beta_q^\infty + \frac{A}{(w-1)^{3/4}} \label{e:betaq}  \\
    &&\beta_p(w) = \beta_p^\infty 
                \left( \frac{w-1}{w+1} \right)^2 \label{e:betap}  
\end{eqnarray}
with 
%
%
\begin{equation}
    \label{e:pinfty}
    p^\infty = \left(
               1 - q^\infty e^{-\beta_q^\infty}
    \right) e^{\beta_p^\infty}
\end{equation}
which overall corresponds to five free fitting parameters
for a global fit of the four functions simultaneously. In practice,
we minimized a cost function given by the sum of the squares of 
the  deviations between the numerical values 
(fitted from the simulations) and interpolating functions~\eqref{e:q},~\eqref{e:p},
~\eqref{e:betaq}~\eqref{e:betap}, summed over the four parameters 
and over the available points on the $w$ axis, $w=[3,5,7,9,11,13]$
(see again Fig.~\ref{fig:Gammapars}). The best-fit values of the 
free parameters are reported in Table~\eqref{tab:Gammapars}.
The corresponding best interpolating functions shown as solid 
lines in Fig.~\ref{fig:Gammapars} demonstrate that a global 
minimum exists that respects the two 
constraints~\eqref{e:limGi} and~\eqref{e:limG1}.

%
%
%
\begin{table}[h!]
    \centering
    \begin{tabular}{c c r}
        \hline
        {\bf Parameter}  & & {\bf Value} \\
        \hline\hline
        $A$              & &  106.75  \\
        $\beta_q^\infty$ & &    3.362 \\
        $\beta_p^\infty$ & &    1.52  \\
        $q^0$            & &    4.77  \\
        $q^\infty$       & &   15.16  \\
        $p^\infty$       & &    2.18  \\
        \hline
    \end{tabular}
    \caption{Best-fit values of the free parameters that enter the 
    interpolating functions~\eqref{e:q},~\eqref{e:p},
    ~\eqref{e:betaq}~\eqref{e:betap} obtained by a simultaneous fit 
    of all four functions on all the available points in $w$ space.}
    \label{tab:Gammapars}
\end{table}



%

\end{document}